\documentclass[A4,aps,twocolumn,superscriptaddress]{revtex4}
\usepackage[dvips]{graphics}
\usepackage{enumerate}

\begin{document}
%%%%%%%%%%%%%%%%%%%%%%%%%%%%%%%%%%%%%%%%%%%%%%%%%%%%%%%%%%%%%%%%%
\preprint{VPI--IPPAP--04--07}
\renewcommand{\thepage}{VPI-IPPAP-04-07}

\title{Phenomenology of Not-so-heavy Neutral Leptons:  
The NuTeV Anomaly, Lepton Universality, and Non-Universal Neutrino-Gauge Couplings}
\author{Tatsu~Takeuchi}%\email{takeuchi@vt.edu}
\affiliation{Institute for Particle Physics and Astrophysics,
Physics Department, Virginia Tech, Blacksburg VA 24061}
\author{Will~Loinaz}%\email{loinaz@alumni.princeton.edu}
\affiliation{Department of Physics, Amherst College, Amherst MA 01002}

\maketitle
%%%%%%%%%%%%%%%%%%%%%%%%%%%%%%%%%%%%%%%%%%%%%%%%%%%%%%%%%%%%%%%%%

The NuTeV experiment \cite{Zeller:2001hh} 
at Fermilab has determined the effective neutrino-nucleon coupling
parameters $g_L^2$ and $g_R^2$ from muon (anti)neutrino-nucleon scattering:
\begin{eqnarray}
g_L^2 & = & 0.30005 \pm 0.00137\;, \cr
g_R^2 & = & 0.03076 \pm 0.00110\;.
\label{nutev}
\end{eqnarray}
The Standard Model (SM) predictions of these parameters 
based on a global fit to non-NuTeV data, cited as 
$[g_L^2]_\mathrm{SM}=0.3042$ and 
$[g_R^2]_\mathrm{SM}=0.0301$ in Ref.~\cite{Zeller:2001hh},
differ from the NuTeV result by $3\sigma$ in $g_L^2$.

The NuTeV value for $g_L^2$ in Eq.~(\ref{nutev}) is 
\textit{smaller} than its SM prediction, reflecting 
the fact that the ratios 
$R_{\nu}=\sigma(\nu_\mu N\!\rightarrow\!\nu_\mu X)/\sigma(\nu_\mu N\!\rightarrow\!\mu^- X)$ and 
$R_{\bar{\nu}}=\sigma(\bar{\nu}_\mu N\!\rightarrow\!\bar{\nu}_\mu X)/\sigma(\bar{\nu}_\mu N\!\rightarrow\!\mu^+ X)$ 
were smaller than expected by the SM. 
Thus, possible \textit{new} physics explanations of the NuTeV anomaly 
would be those that suppress the neutral current cross sections over 
the charged current cross sections, or enhance the charged current 
cross sections over the neutral current cross sections. 
To this end, two classes of models have been devised.

Models of the first class suppress $R_\nu$ and $R_{\bar{\nu}}$ through
new neutrino-quark interactions, mediated by leptoquarks or 
extra $U(1)$ gauge bosons ($Z'$'s), 
which interfere either destructively with the $Z$-exchange amplitude, 
or constructively with the $W$-exchange amplitude \cite{Davidson:2001ji}.
To maintain agreement between the SM and non-NuTeV data,
the new interactions must selectively interfere with the 
$\nu_\mu N$ ($\bar{\nu}_\mu N$) scattering process, but little else.  
This severely restricts the types of interactions that may be introduced.

Models of the second class suppress the $Z\nu\nu$ coupling by mixing the
neutrino with heavy gauge singlet states (neutrissimos,
\textit{i.e.} right-handed neutrinos) \cite{Chang:1994hz,LOTW1,LORTW2}.
For instance, if the $SU(2)_L$ active $\nu$ is a linear combination of two mass eigenstates with
mixing angle $\theta$, 
\begin{equation}
\nu = \nu_\mathrm{light} \cos\theta + \nu_\mathrm{heavy} \sin\theta \;, 
\end{equation}
then
$Z\nu\nu$ is suppressed by  $\cos^2\theta$, and  $W\ell\nu$ is suppressed by $\cos\theta.$
More generally, if the $Z\nu_\ell \nu_\ell$ coupling ($\ell=e,\mu,\tau$) is suppressed by 
a factor of $(1-\varepsilon_\ell)$,  then the $W\ell\nu_\ell$ coupling is suppressed by 
$(1-\varepsilon_\ell/2).$

Such suppressions of the neutrino-gauge couplings affect not only NuTeV
observables. In addition to the suppression of the $Z$ invisible width by a factor of
$[1-(2/3)(\varepsilon_e+\varepsilon_\mu+\varepsilon_\tau)]$, all SM observables will be affected
through the Fermi constant $G_F$ which is no longer equal to the the muon decay constant $G_\mu$:
\begin{equation}
G_F = G_\mu \left(1+\frac{\varepsilon_e + \varepsilon_\mu}{2} \right)\;.
\end{equation}
This shift in $G_F$ would destroy the agreement between the SM and $Z$-pole observables. 
But, since $G_F$ always appears in the combination $\rho\,G_F$ in neutral current amplitudes, 
agreement can be recovered by absorbing the shift in $G_F$ into a shift in $\rho$, 
or equivalently, in the oblique correction parameter $T$~\cite{Peskin:1990zt}.  
The $Z$-pole, NuTeV, and $W$ mass data can all be fit with the oblique correction
parameters $S$, $T$, $U$, and a flavor universal suppression parameter
$\varepsilon = \varepsilon_e = \varepsilon_\mu = \varepsilon_\tau$, with best fit value
$\varepsilon  =  0.0030\pm 0.0010.$~\cite{LOTW1}.

This value of $\varepsilon$  implies a large mixing angle, $\theta = 0.055 \pm 0.010$,  
if interpreted as due to mixing with a single heavy state. 
The traditional seesaw mechanism ties $\theta$ to the ratio of neutrino masses:
\begin{equation}
\frac{m_\mathrm{light}}{m_\mathrm{heavy}}\approx \theta^2\;.
\end{equation}
With $m_\mathrm{light} \sim 0.1\,\mathrm{eV}$ and $m_\mathrm{heavy} \sim 100\,\mathrm{GeV}$ 
(we need $m_\mathrm{heavy}>M_Z$ to suppress $\Gamma_\mathrm{inv}$)
the mixing angle is orders of magnitude too small: $\theta \sim 10^{-6}$. 

By contrast, models with intergenerational mixing have additional degrees of 
freedom which permit them to evade these constraints by decoupling masses and 
mixing angles~\cite{Chang:1994hz,LORTW2}.  
In these models, intergenerational symmetries imposed on the neutrino mass 
texture are the source of the naturally light mass eigenstates.   
No longer fixed at the GUT scale, heavy states might
be relatively light (not-so-heavy), 
the current experimental lower bound being just above the $Z$ mass \cite{L3},
and well within reach of near-future collider experiments.

Large mixing angles between heavy and light states may enhance the rate 
of flavor-changing processes mediated by heavy states.  
Hence, stringent constraints can be placed on these models by limits on lepton flavor violation.
For instance, assuming $m_\mathrm{heavy} \gg M_W$, the MEGA limit on 
$\mu\rightarrow e\gamma$ \cite{mega}  leads to the constraint~\cite{LORTW2}
$\varepsilon_e\varepsilon_\mu <10^{-8}.$ This is clearly incompatible with
$\varepsilon_e=\varepsilon_\mu =0.003$ and implies rather 
$\varepsilon_e \approx 0$ or  $\varepsilon_\mu \approx 0.$  

%%%%%%%%%%%%%%%%%%%%%%%%%%%%%%%%%%%%%%%%%%%%%%%%%%%%%%%%%%%%%%%%%%%%%%%
%\widetext

\begin{figure}[h!]
\begin{center}
\rotatebox{90}{\scalebox{0.38}{\includegraphics{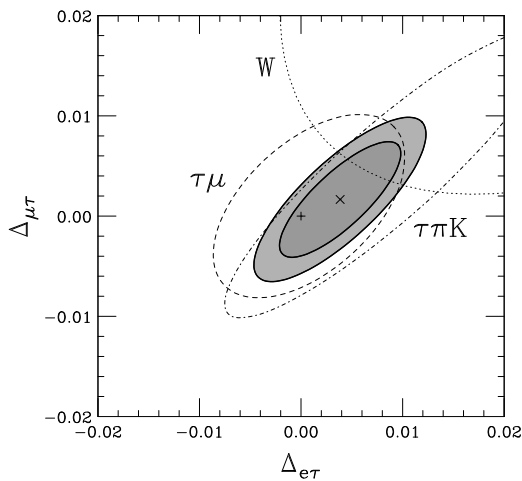}}} 
\caption{The limits on $\Delta_{e\tau}$ and $\Delta_{\mu\tau}$ 
from various decays.
%The shaded areas are the 68\% (dark gray) and 90\% (light gray)
%confidence contours.
}
\label{emufig2}
\end{center}
\end{figure}

%%%%%%%%%%%%%%%%%%%%%%%%%%%%%%%%%%%%%%%%%%%%%%%%%%%%%%%%%%%%%%%%%%%%%%%
%%%%%%%%%%%%%%%%%%%%%%%%%%%%%%%%%%%%%%%%%%%%%%%%%%%%%%%%%%%%%%%%%%%%%%%
\begin{figure}[h!]
\begin{center}
\rotatebox{90}{\scalebox{0.38}{\includegraphics{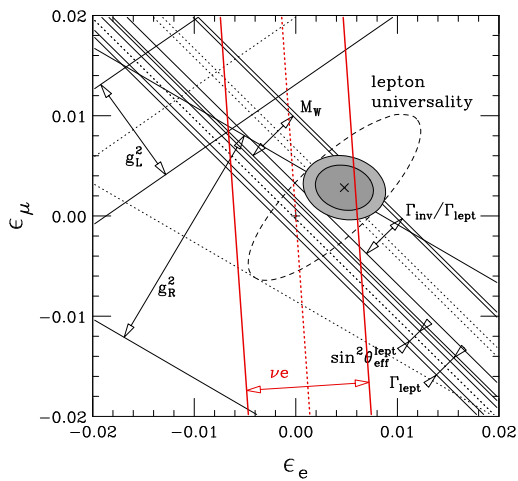}}}
\caption{Limits on $\varepsilon_e$ and $\varepsilon_\mu$.
The $1\sigma$ bands are shown for each observable ignoring correlations.
The shaded areas are the 68\% (dark gray) and 90\% (light gray)
confidence contours including correlations.
The red band illustrates the potential impact of a 
proposed reactor measurement of the $\bar{\nu}_e e$ cross section.}
\label{emufig}
\end{center}
\end{figure}

%%%%%%%%%%%%%%%%%%%%%%%%%%%%%%%%%%%%%%%%%%%%%%%%%%%%%%%%%%%%%%%%%%%%%%%
%\narrowtext

Such a pattern of $\varepsilon_\ell$ will generically induce
violations of lepton universality in charged-current processes.  
The fit to lepton universality violating parameters 
$\Delta_{e \tau} \equiv \varepsilon_e - \varepsilon_\tau$ and 
$\Delta_{\mu \tau} \equiv \varepsilon_\mu - \varepsilon_\tau$ to data from 
$W$, $K$, $\pi$, $\tau$, and $\mu$ decays is shown
in Fig.~\ref{emufig2}.  Best-fit values are:
\begin{eqnarray}
\Delta_{e \tau}&=&0.0039 \pm 0.0040 \;,\cr
\Delta_{\mu \tau}&=&0.0017 \pm 0.0038\; .
\end{eqnarray}
Unfortunately, the quality of the fit is unimpressive 
($\chi^2/\text{d.o.f.} = 8.4/5$). 
In addition, the internal consistency of the data determining some of the 
branching ratios entering the fit, especially $B(\tau \rightarrow \pi \nu_\tau)$, 
is poor.  Thus, definitive bounds on lepton universality await additional data.

Fits to $\varepsilon_\ell$ and $S$,$T$,$U$ using both electroweak data and the 
best-available lepton universality constraints indicate that the data are 
compatible with several patterns of neutrino-gauge coupling suppression.  
The result of a five-parameter fit to $S$,$T$,$U$,$\varepsilon_e$, and 
$\varepsilon_\mu$ is shown in Fig.~\ref{emufig}~\cite{Loinaz:2004qc}.  
Among the models analyzed in Ref.~\cite{Loinaz:2004qc}, the model with
$\varepsilon_e \neq 0$, $\varepsilon_\mu = \varepsilon_\tau = 0$ best 
accommodates both the fit data and the MEGA constraint, with best-fit values
\begin{eqnarray}
S & = & -0.04 \pm 0.10 \;,\cr
T & = & -0.46\pm 0.15 \;,\cr
U & = & \phantom{-}0.52\pm 0.16  \;,\cr
\varepsilon_e & = & 0.0051\pm 0.001\;,
\end{eqnarray}
for a reference SM with $m_H=115\;\mathrm{GeV}.$

Further experimental constraints to neutrino-neutrissimo
mixing models are expected in the near future.
For instance, a proposed improved reactor measurement of the ${\bar\nu}_e e$ 
cross section would provide a very clean and direct measurement of 
$\varepsilon_e$ \cite{Conrad:2004gw} and could significantly improve the current bound.  
The potential bound from a measurement at the $1.3\%$ level is illustrated 
by the red band in Fig.~\ref{emufig}, and 
even more precise measurements are contemplated.
The consistency of the charged-current lepton universality constraints 
from various decays is currently poor but should improve with additional data.  
Searches for lepton flavor violation at MEG, MECO, and elsewhere are underway.

%%%%%%%%%%%%%%%%%%%%%%%%%%%%%%%%%%%%%%%%%%%%%%%%%%%%%%%%%%%%%%%%%%%%%%%
\section*{Acknowledgments}

This paper was presented by Takeuchi at the 
YITP workshop `Progress in Particle Physics' 2004.
We would like to thank Peter Fisher, Naotoshi Okamura, Saifuddin Rayyan, and L.C.R. Wijewardhana for their collaboration on this project.  
This research was supported in part (T.T.) by the U.S. Department of Energy,
grant DE-FG05-92ER40709, Task A.

%%%%%%%%%%%%%%%%%%%%%%%%%%%%%%%%%%%%%%%%%%%%%%%%%%%%%%%%%%%%%%%%%%%%%%%

%%%%%%%%%%%%%%%%%%%%%%%%%%%%%%%%%%%%%%%%%%%%%%%%%%%
%%%%%%%%%%%%%%%%%%%%%%%%%%%%%%%%%%%%%%%%%%%%%%%%%%%%%%%%%%%%%%%%%%
\end{document}